\documentclass{article}

\usepackage{microtype}
\usepackage{graphicx}
\usepackage{booktabs} 
\usepackage[caption=false]{subfig}

\usepackage{hyperref}

\usepackage[accepted]{icml2024}

\usepackage{amsmath}
\usepackage{amssymb}
\usepackage{mathtools}
\usepackage{amsthm}

\usepackage[capitalize,noabbrev]{cleveref}

\definecolor{red}{rgb}{1,0,0}

\theoremstyle{plain}

\theoremstyle{definition}

\theoremstyle{remark}

\usepackage[textsize=tiny]{todonotes}

\icmltitlerunning{Submission and Formatting Instructions for ICML 2024}

\begin{document}

\twocolumn[
\icmltitle{Variational and Explanatory Neural Networks for Encoding Cancer Profiles and Predicting Drug Responses}



\icmlsetsymbol{equal}{*}

\begin{icmlauthorlist}
\icmlauthor{Tianshu Feng}{gmu-sero,mayo}
\icmlauthor{Rohan Gnanaolivu}{mayo}
\icmlauthor{Abolfazl Safikhani}{gmu-stat}
\icmlauthor{Yuanhang Liu}{mayo}
\icmlauthor{Jun Jiang}{mayo}
\icmlauthor{Nicholas Chia}{anl}
\icmlauthor{Alexander Partin}{anl}
\icmlauthor{Priyanka Vasanthakumari}{anl}
\icmlauthor{Yitan Zhu}{anl}
\icmlauthor{Chen Wang}{mayo}
\end{icmlauthorlist}

\icmlaffiliation{gmu-sero}{Department of Systems Engineering and Operations Research, George Mason University, Fairfax, VA, USA}
\icmlaffiliation{gmu-stat}{Department of Statistics, George Mason University, Fairfax, VA, USA}
\icmlaffiliation{mayo}{Department of Quantitative Health Sciences, Mayo Clinic, Rochester, MN, USA}
\icmlaffiliation{anl}{Division of Data Science and Learning, Argonne National Laboratory, Lemont, IL, United States}

\icmlcorrespondingauthor{Tianshu Feng}{tfeng@gmu.edu}
\icmlcorrespondingauthor{Chen Wang}{Wang.Chen@mayo.edu}

\icmlkeywords{Human cancers, transcriptomics, variational inference, neural network, drug response prediction, local interpretability}

\vskip 0.3in
]



\printAffiliationsAndNotice{}  

\begin{abstract}
Human cancers present a significant public health challenge and require the discovery of novel drugs through translational research. Transcriptomics profiling data that describes molecular activities in tumors and cancer cell lines are widely utilized for predicting anti-cancer drug responses.  However, existing AI models face challenges due to noise in transcriptomics data and lack of biological interpretability.  To overcome these limitations, we introduce VETE (Variational and Explanatory Transcriptomics Encoder), a novel neural network framework that incorporates a variational component to mitigate noise effects and integrates traceable gene ontology into the neural network architecture for encoding cancer transcriptomics data.  Key innovations include a local interpretability-guided method for identifying ontology paths, a visualization tool to elucidate biological mechanisms of drug responses, and the application of centralized large scale hyperparameter optimization. VETE demonstrated robust accuracy in cancer cell line classification and drug response prediction. Additionally, it provided traceable biological explanations for both tasks and offers insights into the mechanisms underlying its predictions. VETE bridges the gap between AI-driven predictions and biologically meaningful insights in cancer research, which represents a promising advancement in the field.
\end{abstract}

\section{Introduction} \label{sec:intro}

Cancer is a significant global health issue that requires a comprehensive spectrum of anti-cancer drug research, including the discovery of novel compounds, biochemical optimizations, preclinical evaluations, and the final translational stages of rigorous clinical trials \citep{gutierrez_next_2009, cui_discovering_2020, honkala_harnessing_2022}. In recent years, integrating AI approaches into molecular data and anti-cancer drug response studies has revolutionized our understanding and capabilities in the field of cancer research \citep{albaradei_machine_2021, franco_performance_2021, leng_benchmark_2022, partin2023deep}. These advanced computational methods allow researchers to unravel complex biological networks and pathways with unprecedented precision and depth. In data-driven omics characterization of molecular and biology profiles, deep learning algorithms have been instrumental in decoding large and intricate datasets, ranging from genomics to proteomics, thus enhancing our understanding of cellular mechanisms and disease pathogenesis \citep{zou_primer_2019, aliper_deep_2016, bulik-sullivan_deep_2019}. In drug response studies, deep learning has shown promise in predicting drug efficacy and toxicity, personalizing oncology approaches, and accelerating drug discovery processes \citep{chang_cancer_2018, kuenzi_predicting_2020}. The ongoing advancements in AI algorithms, neural network architecture, and computational power continue to open new frontiers in cancer research, promising transformative impacts on therapeutic development and precision oncology. 

Despite the advancements brought by deep learning in cancer omics and drug response research, these multilayered neural models are limited by their lack of transparency in understanding biological mechanisms. Often operating as “black boxes,” deep learning models provide limited insight into how they derive their conclusions. This opacity is a critical drawback in biomedical research, where understanding the biological underpinnings is as crucial as predictive accuracy for formulating testable hypothesis and determining follow-up biological experiments. For instance, in omics studies, while deep learning models can efficiently predict cell types based on gene expression patterns or genetic variants  \citep{dong_semi-supervised_2022}, they often fail to  offer clear explanations for these predictions, which is vital for clinical decision-making and therapeutic development  \citep{haywood_pam50_2022, kircher_assessing_2023, pleasance_whole-genome_2022}. Similarly, in drug response prediction, although these models can identify potential responders and non-responders to a treatment, their inability to explain the basis of these predictions limits their clinical applicability and translational research aimed at overcoming drug resistance \citep{adam_machine_2020, baptista_deep_2021}. These challenges highlight the need for more interpretable and explainable AI models in biomedical research, where the understanding of 'why' and 'how' a model makes a prediction is just as important as the prediction itself \citep{rudin_stop_2019}.

\begin{figure*}[!htb]
    \centering
    \subfloat[\label{fig:overview}]{\includegraphics[width=.75\textwidth]{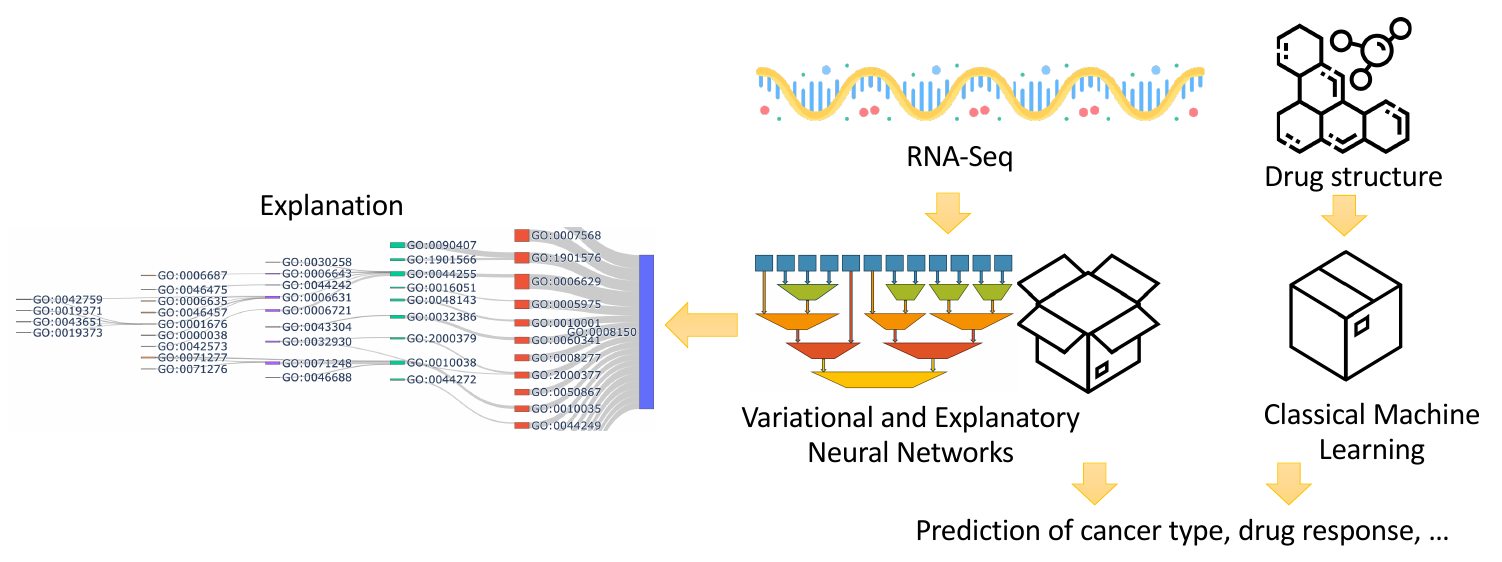}}\\\vspace{-4ex}
  	\subfloat[\label{fig:vae}]{\includegraphics[width=.25\textwidth]{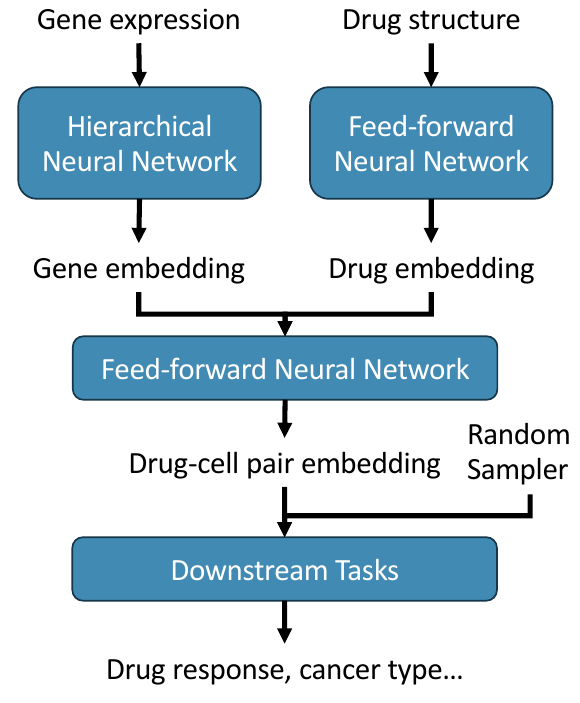}}\quad\quad\quad
	\subfloat[\label{fig:gene_embedding}]{\includegraphics[width=.45\textwidth]{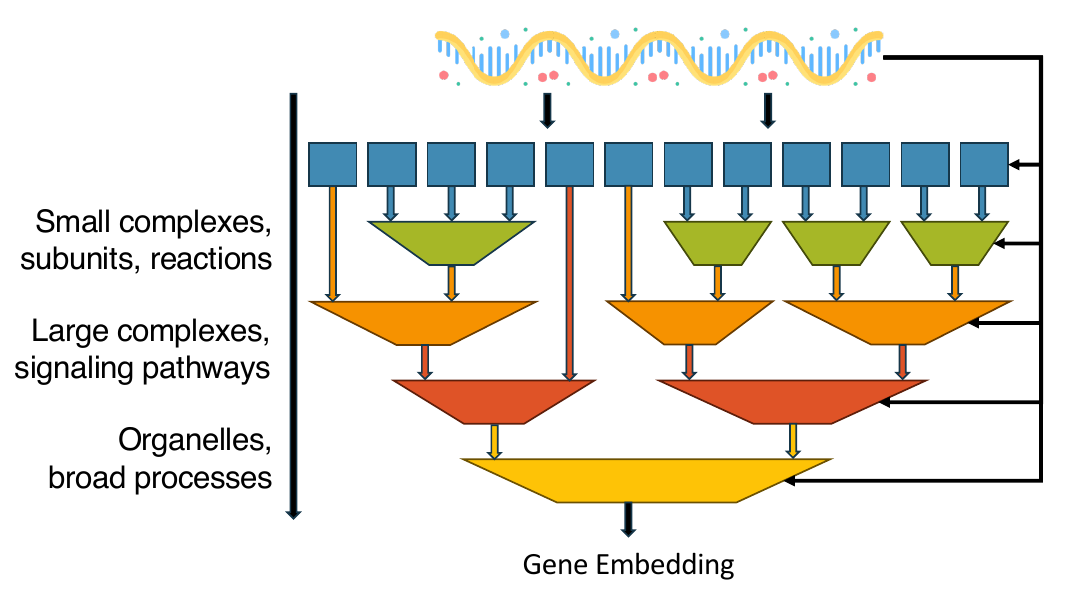}}\\
    \caption{Model design: (a) Overall VETE framework; (b) Model structure for drug-cell embedding and downstream tasks; (c) Hierarchical neural network (HNN) following the hierarchical structure of an ontology of cellular subsystems to encode gene expression of samples.}
    \label{fig:model_design}
\end{figure*}

In this paper, we propose the Variational and Explanatory Transcriptomics Encoder (VETE), based on variational information bottleneck principle \citep{alemi2016deep}. VETE can be configured with various deep learning components (hereby referred as submodel), including a submodel for encoding cancer transcriptomics data, and can incorperate additional submodels for extracting embeddings from other omics data  and/or drug data (Figure \ref{fig:overview}). The transcriptomics submodel’s neural network structure is designed according to the Gene Ontology, a literature-curated knowledge database encompassing various aspects of biology subsystems (intracellular components, processes or functions) \citep{gene_ontology_consortium_expansion_2017}. Each parent-child relationship between these subsystems is represented as a connection in the network, forming a hierarchical model structure that explicitly models the biological process at different scales and links gene expression to molecular and functional relevance. The drug embedding submodel, on the other hand, is based on multi-layer perceptron (MLP) with drug fingerprint as input and drug molecular structure embedding as output \citep{kuenzi_predicting_2020}.The outputs of these submodels are concatenated to generate latent random variables, from which latent realizations are drawn for downstream tasks, such as cancer type classification and drug response prediction. Previous studies have used the Gene Ontology to predict phenotypes with disrupted genes in a eukaryotic cell, albeit using binary genetic mutation states as features \citep{kuenzi_predicting_2020, ma_using_2018}. Despite the hierarchical structure, understanding the model decision process for individual predictions remains challenging. In this paper, we propose a graph-based model explanation technique to reveal the influence of subsystems on final model predictions, leading to a more transparent approach. 

The model is validated with two datasets: Genomics of Drug Sensitivity in Cancer (GDSC) \citep{yang_genomics_2012} and The Cancer Genome Atlas Program (TCGA) \citep{weinstein_cancer_2013} (detailed in the appendix). We evaluated the model performance in cancer type classification detailed the transcriptomics submodel on both datasets and further study the drug response prediction using GDSC.  The results demonstrate that VETE efficiently identifies key drug-cell interaction factors consistent with existing literature, even without explicitly incorporating prior biological knowledge during model training. These findings suggest VETE's substantial potential to advance drug-cell interaction studies and enhance our understanding of biological mechanisms.

\section{Methodology}
\subsection{Data Preparation}
For model training and analysis, we selected the top 15\% most frequently mutated genes in human cancers, as identified in the Cancer Cell Line Encyclopedia (CCLE) \citep{barretina_cancer_2012}. We focused on genes annotated with Gene Ontology (GO) terms \citep{ashburner_gene_2000}, resulting in a set of 3,008 genes. GO terms were retained only if they included at least 10 of these selected genes and were distinct from all their child terms. To reduce model complexity, we further limited the hierarchy to a maximum depth of five. The final hierarchy consists of 2,086 GO terms and defines the structure of the VETE submodel for encoding cancer transcriptomics data.

Multiple datasets were utilized for model training and validation. Multi-omics data of cell lines, including gene expressions and cancer types,  were extracted from the Dependency Map (DepMap) portal of CCLE. Drug information was retrieved from PubChem \citep{kim_pubchem_2016}. Using the SMILES (Simplified Molecular Input Line Entry Specification) nomenclature for drugs  \citep{weininger_smiles_1988}, we calculated molecular fingerprints and descriptors to embed drugs' molecular structures using the Mordred \citep{moriwaki_mordred_2018} and RDKit \citep{landrum_rdkit_2006} Python packages. Cell line response data were extracted from GDSC version 2 \citep{yang_genomics_2012}. In total, we included 66,353 unique drug-cell line pairs with 1,007 cell line samples and 1,565 drugs. 

\subsection{Variational and Explanatory Transcriptomics Encoder for Biological Graphs} \label{sec:VETE}

The proposed Variational and Explanatory Transcriptomics Encoder (VETE) consists of expandable submodels for learning the latent embeddings of gene expressions and drug molecular structures  (Figure \ref{fig:vae}). The submodel for transcriptomics encoding is based on a hierarchical neural network (HNN) \citep{ma_using_2018} (Figure \ref{fig:gene_embedding}). This HNN architecture mirrors the hierarchical structure of an ontology of cellular subsystems, where each subsystem’s parent-child relation is modeled with an MLP. The drug embedding submodel encodes the drug structure using a fully connected feed-forward neural network. VETE follows the variational information bottleneck (VIB) framework \citep{alemi2016deep}, where the drug-cell pair embedding is represented as a latent distribution rather than a deterministic vector. Compared with classic autoencoder structure, theoretical and practical studies suggest that VIB framework is more robust against data perturbations and noises \citep{kingma_auto-encoding_2014, alemi2016deep, camuto_towards_2021, korshunova_closer_2021}. This robustness is beneficial given the relatively large number of parameters (over 180 million) and the pre-specified sparse model structure in VETE, as well as the noisy gene expression data with a limited sample size (114,000 drug-cell pairs from GDSC version 2).

We consider a general directed acyclic biological graph $G=(V,E)$ with $M^{V}$ biological terms as nodes $V=\{1,\ldots,M^{V}\}$ (e.g., genes, pathways, or biological processes) and their directed association as edges $E = \{i \rightarrow j | i,j \in V\}$ (e.g., gene $i$ directly influences a biological process $j$). The graph can be curated and constructed based on existing literature and findings, including the Gene Ontology (GO) \citep{gene_ontology_consortium_expansion_2017}, Clique-eXtracted Ontology (CliXO) \citep{dutkowski2013gene}, and kyoto encyclopedia of genes and genomes (KEGG) \citep{kanehisa2000kegg, chanumolu2021kegg2net}.

Let the samples be  $(x^{g,1},x^{d,1},y^1 ), \ldots$, $(x^{g,N},x^{d,N}, y^N )$, where gene feature vector $x^{g,n}$ represents the vector of gene expression values, drug feature vector $x^{d,n}$ denotes the structure embedding of a drug, and $y^n$ is the response variable of the $n$-th drug-cell pair. For the gene expression embedding submodel, 
we consider a general model framework for directed biological graph. First, each node $i$ is represented with an embedding vector $v_i$. Then, for a given node $j$ and and its parents $E_{.j}=\{i \in E|i\rightarrow j \}$, the relationship of $E_{.j}$ and $j$ can be estimated with an almost everywhere differentiable function $f_{.j}$ using the training set such that $v_j \approx f_{.j}(v_i, i \in E_{.j})$. 
In this paper, we model the relationship of $\{i| i \in E_{.j}\}$ and $j$ using an MLP, where the $M_{.j}$ dimensional embedding of the subsystem, $L_{.j} \in \mathbb{R}^{M_{.j}}$, is defined as:
\begin{equation}
    L_{.j} = \textrm{BatchNorm}\left( \tanh (\textrm{MLP}(I_{.j}))\right), \label{eq:MLP}
\end{equation}
where $I^s = \left(L_{.i}, i \in E_{.j}; x^{g}_{.j}\right)$ is the concatenation of $L_{.i}$ for $i \in E_{.j}$ and gene expression features $x^{g}_{.j}$ directly related to node $j$. Here, $\tanh$ represents the nonlinear hyperbolic tangent transformation, and BatchNorm regularizes the model weights to mitigate internal covariate shift and smooth the objective function \citep{santurkar_how_2018}. The gene expression embedding from the last layer of the submodel is denoted as $L^g$. 
The drug embedding submodel follows a classic feed-forward neural network structure with three layers, where the input is the 512-bit ECFP4 fingerprints of drugs \citep{rogers_extended-connectivity_2010}. The output of the final layer, $L^d,$ represents the drug embedding learned by the model. 

The combined layers $L^g$ and $L^d$ are input into an additional fully connected layer. This layer outputs two parameter vectors: $\mu$ and $\sigma$, both in $\mathbb{R}^M$, where $M$ is the dimension of the latent space. These vectors represent the mean and standard deviation of the latent random variable distribution $Z=(Z_1,\ldots,Z_M)\in\mathbb{R}^M$ for the drug-cell pairs, which can be considered as the  encoded latent representation. For downstream applications such as classification or regression, we employ decoder models $D$, where $D$ can be a classifier, regressor, or a neural network designed to reconstruct the input from low-dimensional embeddings. These models are tailored to the specific response variable $y$, using realizations $z\sim Z$ as their input. Notably, $y$ can represent various biological responses, such as the Area Under the dose-response Curve (AUC) for drug response or different cell types. Since the decoder $D$ is jointly trained with the encoder, it can serve both as an auxiliary model during the training stage and as a predictive model during the inference stage.

We consider the mean-field assumption such that elements in $Z$ are independent of each other and follows a normal distribution. Then the objective function of the proposed model is:
\begin{align}
    \frac{1}{N} & \sum_{n=1}^N  E_{z^n\sim Z^n } Loss(D(z^n ),\ y^n) \notag\\ 
    &-\beta\ KL(q(z^n | x^{d, n},x^{g, n}\ )||p(z^n)) \\ &+\lambda\sum_{j} Loss(\textrm{MLP}(L^n_{.j} ), y^n ) .\notag
\end{align}
Here, $Loss$ is the loss function comparing the output of the model and the response variable, $KL(q(.)||p(.))$ is the KL divergence between two distributions with probability density functions $q$ and $p$, $q(z | x^d,x^g\ )$ is the posterior distribution of $Z$ given input pair $(x^d,x^g)$, $p(z)$ is the pre-specified prior of $Z$, and $\beta$ and $\lambda$ are regularization parameters balancing the influence of KL divergence and prediction performance of subsystems. Considering the relatively large scale of the network, following the idea of GoogLeNet \citep{szegedy_going_2015}, we include auxiliary classifiers connected to the individual GO terms in the objective function with a discount weight $\lambda$. In this way, individual GO terms are encouraged to contribute to the overall model performance, and gradients can be propagated back through layers more effectively.

\subsection{Identifying Critical Hierarchical Paths with Local Model Explanation Techniques}

While HNN is considered due to its architecture reflecting the hierarchical organization of cellular subsystem's GO terms, the contribution of each subsystem to the model's output for a specific sample through the ontology graph is not immediately clear. In this paper, we propose graph integrated  gradients (GIG), a local post-hoc model explanation method designed to assign importance scores to the parent-child relationship of subsystems within a biological graph, following the idea of integrated gradients \citep{sundararajan_axiomatic_2017}. 
As a local explanation method, instead of explaining the overall estimated model, GIG focuses on explaining how model makes decisions for individual samples. This local focus is particularly advantageous for drug responses prediction tasks, given the diverse origins of cells from various tumor and organ types. Local explanations enable the identification of specific GO terms contributing to the drug responses of drug-cell pairs pertinent to particular cell or drug  types.  

Specifically, for a model $f$, a point of interest $X$, and a baseline point $X'$, the proposed IGNIM for an edge $i \rightarrow j$ within a directed biological graph is defined as
\begin{align}\label{eq:ignim_dag}
    & \textrm{GIG}_{i\rightarrow j} \left( v_{.j}; v_{.j}^\prime, \Lambda, Q \right) \\
    = & ( v_i - v_i^\prime ) \sum_{k=1}^{m} \lambda_k \left( \int_{0}^1 \frac{\partial \widehat{f}_{.j}(v_{.j}^\prime + \alpha(v_{.j}-v_{.j}^\prime))}{\partial v_i} d\alpha \right),\notag
\end{align}
where $v_{.j}$ is the vector of all $v_i$s for $i \in E_{.j}$. In VETE, we let $v_{.j} = L_{.j}$, $v_i=L_{.i}$, and $f$ as defined in (\ref{eq:MLP}). 

The point of interest in our GIG explanation method refers to the specific sample or group of samples for which we aim to explain the model's prediction mechanism. This could be a particular drug-cell pair or a set of pairs from the same tumor type. For groups, we average the GIG scores across all points to derive a collective explanation. 

The baseline represents the reference point or population against which comparisons are made, and the choice of which significantly affects the explanation's outcome. The selection of an appropriate baseline can be based on the fitted model and dataset, such as the model's decision boundary or the dataset's expectation \citep{nair_explaining_2022}. To enhance interpretability, context-specific baselines can be employed. For instance, to understand why a model predicts drug DTX to be more effective for BRCA than LUAD, DTX-BRCA pairs can serve as the point of interest, with DTX-LUAD pairs as the baseline. Through GIG, we can then uncover the features most influential in the sensitivity prediction difference between BRCA and LUAD. For simplification and demonstration purpose, in this paper, unless specified otherwise, we use zero values as the baseline, representing scenarios where gene expression is unobserved.

\subsection{Visualization with Sankey Plot}\label{sec:sankey}

\begin{algorithm}
\caption{Searching and Pruning Algorithm}\label{alg:prune}
\begin{algorithmic}

\STATE {\bfseries Input:} $G=(V,E)$, $\textrm{GIG}_{i\rightarrow j}$ for $ij\in E$, $p$ the number of nodes to be kept.

\STATE Initialize $S \leftarrow \{S_0\}$.
\WHILE{$\{i \in G_S|i \rightarrow j, j\in G_T\} \neq \Phi $}
    \FOR{$j$ in $G_T$}
\STATE Select nodes $i$ from $G_S$ with the highest $|\textrm{GIG}_{i\rightarrow j}|$
\STATE Add selected nodes and associated edges to $G_T$
\ENDFOR
\ENDWHILE
\end{algorithmic}
\end{algorithm}

We visualize GIG scores using a Sankey plot to aid in interpretation, where nodes represent the subsystems, and edges represent the flows of importance scores from the subsystems to the final model output through the hierarchical oncology graph. 

However, The complexity of Sankey plots, often containing hundreds to thousands of GO terms and parent-child pairs, raises a challenge for effective visualization and interpretation. We develop a searching and pruning algorithm to identify and highlight the most important paths. 
With an empty graph $G_T$, the process begins by adding the sink node $M^V$ from the original Sankey plot $G_S$ to $G_T$. Subsequently, we consider a recursive process starting from the second-highest level of $G_S$. In each iteration, we identify and add to $G_T$ the nodes in $\{i \in G_S|i \rightarrow j, j\in G_T\}$ with the highest absolute GIG scores. This searching procedure is repeated until $\{i \in G_S|i \rightarrow j, j\in G_T\}$ is empty. The pseudo code of the process can be found in Algorithm \ref{alg:prune}.

\subsection{Large Scale Hyper Parameter Optimization} \label{sec:HPO}

Model explanation results are influenced by both data and the fitted model \citep{chen2020true}. Although VETE combines interpretable model structures and explanation methods to uncover drug-cell interaction mechanisms in the proposed framework, the explanations can still be inaccurate if model fitness is suboptimal \citep{molnar2020general}. Issues, such as overfitting, underfitting, and poor convergence can lead to less valuable explanations as they reflect faulty model behaviors. Therefore, gaining accurate insights into drug-cell interactions and drug response predictions require well-fitted models. 

Given the numerous hyperparameters involved in VETE, hyperparameter optimization (HPO) is critical for improving the model prediction and explanation. In this study, we utilize DeepHyper \citep{balaprakash2018deephyper}, an asynchronous hyperparameter search package, which offers parallel implementations of the SMAC (Sequential Model-Based Algorithm Configuration) algorithm \citep{hutter2011sequential}, enabling HPO on large scale high-performance computers. 

We specifically used Bayesian optimization (BO) \citep{jones1998efficient}, known for its efficiency in global optimization through minimizing the number of direct queries to the real “expensive” black-box function by iteratively updating an internal surrogate model. Unlike grid search and random search, BO effectively explores the high-dimensional hyperparameter space by balancing exploration and exploitation \citep{wu2019hyperparameter}. To fully utilize the parallel capacity of HPC, DeepHyper parallelizes BO through a centralized architecture, where a central coordinator assigns hyperparameter evaluations to remote workers. We use Extremely Randomized Trees \citep{geurts2006extremely} as the surrogate model for its improved epistemic uncertainty estimates from a ran-dom-split strategy in tree construction. By leveraging HPO on large-scale HPC, we aim for the VETE model to accurately learn intrinsic drug-cell interactions and provide reliable drug response predictions and explanations.

\section{Experiments}

\subsection{Model Training Tasks}

\begin{table*}[htbp]
  \centering
  \caption{Hyperparameter optimization setup and final values.}\label{tab:HPO_params}%
   \resizebox{0.8\linewidth}{!}
    {
    \begin{tabular}{llccc}
    \toprule
     Description & Prior & Range & \multicolumn{1}{l}{Default value} & \multicolumn{1}{l}{Final value} \\
    \midrule
      Number of epochs & Uniform & [2, 40] & 10    & 29 \\
      Batch size & Log uniform & [64, 2048] & 256   & 449 \\
      Learning rate of Adam & Log-uniform & [0.0001, 0.01] & 0.001 & 5.60E-03 \\
      Dimension of gene embedding, $M^g$ & Uniform & [2, 10] & 6     & 4 \\
      Dimension of latent space, $M$ & Uniform & [2, 10] & 6     & 6 \\
      $\lambda$ for balancing subsystems & Uniform & [0.1, 0.5] & 0.2   & 0.33 \\
      Epsilon of ADAM optimizer & Uniform & [0.000001, 0.0001] & 1.00E-05 & 3.60E-05 \\
     $\beta$ for $\beta$-VAE & Log-uniform & [0.0001, 1] & 0.001 & 2.20E-03 \\
    \bottomrule
    \end{tabular}%
    }
\end{table*}%

\textbf{Top-5 cancer type classification (cell line tumors from GDSC).} To assess the transcriptomics submodel’s effectiveness for gene expression embedding in cancer cell lines, we conducted a classification task. The dataset comprised transcriptomics data from cancer cells, divided into 806 training and 201 testing samples. Given the dataset's imbalanced observations across 138 cancer types, we reframed the problem as a binary classification task. The objective was to identify whether a sample belonged to one of the five most common cancer types (LUAD (lung adenocarcinoma), COAD (colon adenocarcinoma), SCLC (small cell lung cancer), GB (gallbladder cancer), PAAD (pancreatic adenocarcinoma)). For this task, we implemented an MLP with one hidden layer and a singular output neuron as the decoder. This design choice was intended to improve the encoder's learning of sample embeddings. Notably, the drug embedding submodel was excluded from this task. We used binary cross-entropy as the loss function for evaluating model performance.

\textbf{Drug response prediction (GDSC).} We applied VETE to predict drug response. Our dataset includes drug-cell pairs from GDSC 2, split into 95,373 training samples and 19,271 testing samples. We ensured no overlap of cell lines between the training and testing sets. The inputs are drug-cell pairs, and the response variable is the Area Under the dose-response Curve (AUC) for drug responses. Similar to the classification task, we used an MLP with one hidden layer as the decoder to promote effective learning of sample embeddings by the encoder. The loss function for this task is mean squared error, aligning with the regression nature of the task.

We applied VETE to drug response prediction in two steps: first evaluating the transcriptomics submodel's effectiveness in encoding gene expression through a binary classification task, and then incorporating drug embedding for drug response prediction. We used MLP, XGBoost, and Random Forest as baselines for comparison in both tasks. The MLP has 3 hidden layers with 64, 32, and 16 neurons and ReLU activation function, and was trained with Adam. XGBoost employed 100 gradient boosted trees with a maximum depth of 4. Random Forest has 100 trees in the forest.

\subsection{VETE for Cell Line Classification} \label{sec: top-5_clf}

\begin{figure}[!tb]
    \centering
    \subfloat[\label{fig:top-5-ROC}]{\includegraphics[width=.245\textwidth]{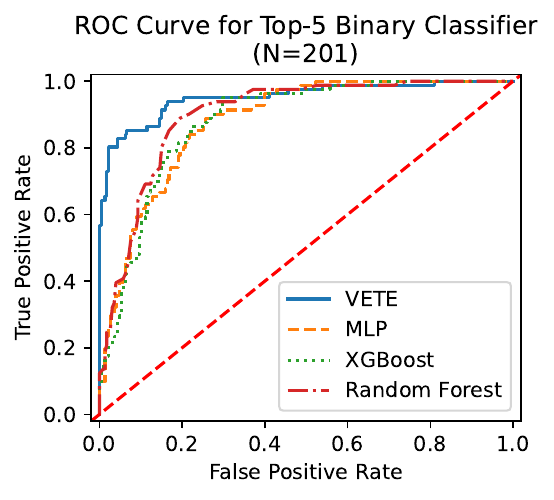}}
  	\subfloat[\label{fig:top5_compare}]{\includegraphics[width=.235\textwidth]{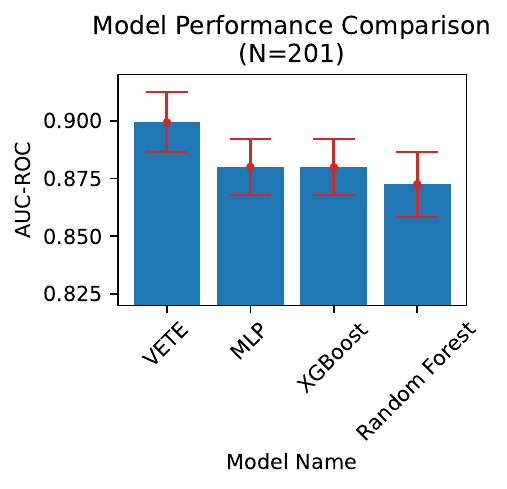}}\\
    \caption{Cell line classification results: (a) ROC curve for VETE, MLP, XGBoost, and Random Forest; (b) Comparison of VETE, MLP, XGBoost, and Random Forest using AUC-ROC.}\vspace{-3ex}
    \label{fig:top-5}
\end{figure}


We trained the VETE model to differentiate between the five most common cancer types (LUAD, COAD, SCLC, GB, PAAD) and other types, using gene expression data as input. The optimization was performed using stochastic gradient descent. Evaluation on the testing set focused on the Receiver Operating Characteristic (ROC) curve and the Area Under the Receiver Operating Characteristic Curve (AUC-ROC). Hyperparameters related to network structure and optimization were crucial for VETE’s performance. We searched for the optimal set of hyperparameters, which are listed in Table \ref{tab:HPO_params}. Model initialization followed the methods introduced in \citep{he_delving_2015}, and optimization was carried out using Adam.  The model is implemented in PyTorch 1.13.1 on Tesla V100 GPUs. For each sample, we took 20 realizations and used the average logits as the final output.

As shown in Figure \ref{fig:top-5}, VETE demonstrates a notable advantage in identifying the top five cancers compared to MLP, XGBoost, and Random Forest. The ROC and AUC-ROC metrics indicate that while MLP, XGBoost, and Random Forest perform comparably, VETE outperforms them. The advantage in performance implies the effectiveness of VETE in learning the gene embedding of cell line samples, which suggests the usefulness of the transcriptomics submodel for the subsequent task of drug response prediction.


\subsection{VETE for Drug Response Prediction}

\begin{figure}[!tb]
    \centering
    \subfloat[\label{fig:drugcell_perf_box}]{\includegraphics[width=.242\textwidth]{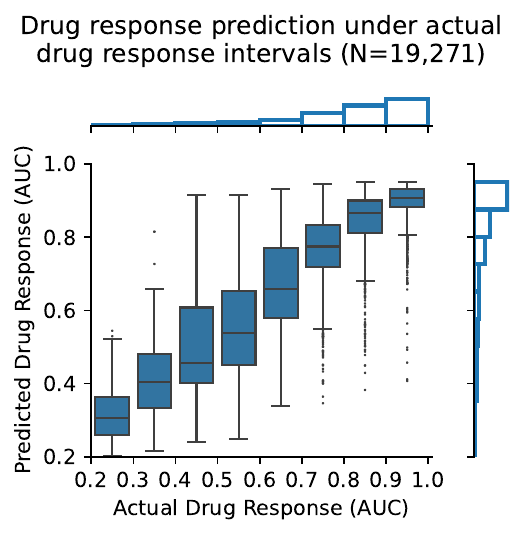}}
  	\subfloat[\label{fig:drugcell_compare}]{\includegraphics[width=.244\textwidth]{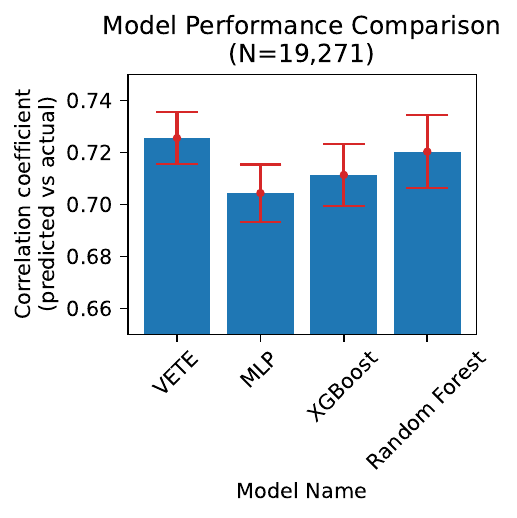}}\\
    \caption{Drug response prediction results: (a) predicted vs. actual drug responses across all drug-cell pairs; (b) performance comparison of VETE again MLP, XGBoost, and Random Forest.}\vspace{-2ex}
    \label{fig:drug_cell}
\end{figure}

\begin{figure}[!tb]
    \centering
    \includegraphics[width=.425\textwidth]{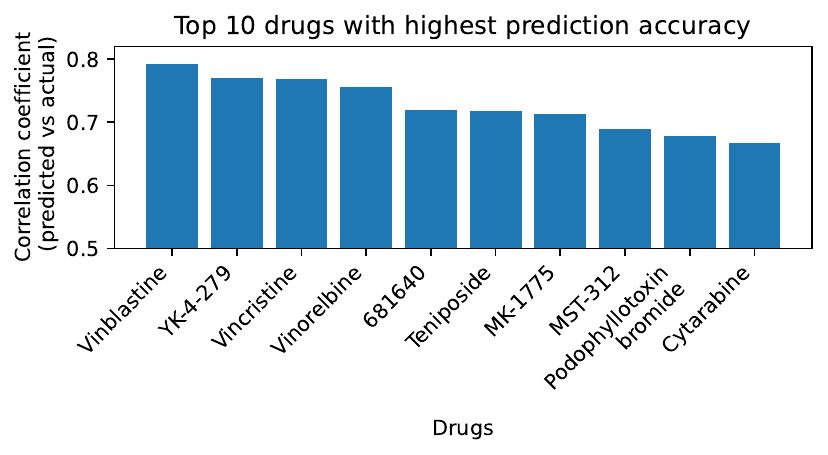}\vspace{-3ex}
    \caption{Top 10 drugs with the highest drug response prediction accuracy.}\vspace{-3ex}
    \label{fig:indi_drug}
\end{figure}

\begin{figure*}[!htb]
    \centering
	\subfloat[\label{fig:sankey_DOX_OV}]{\includegraphics[width=.92\textwidth]{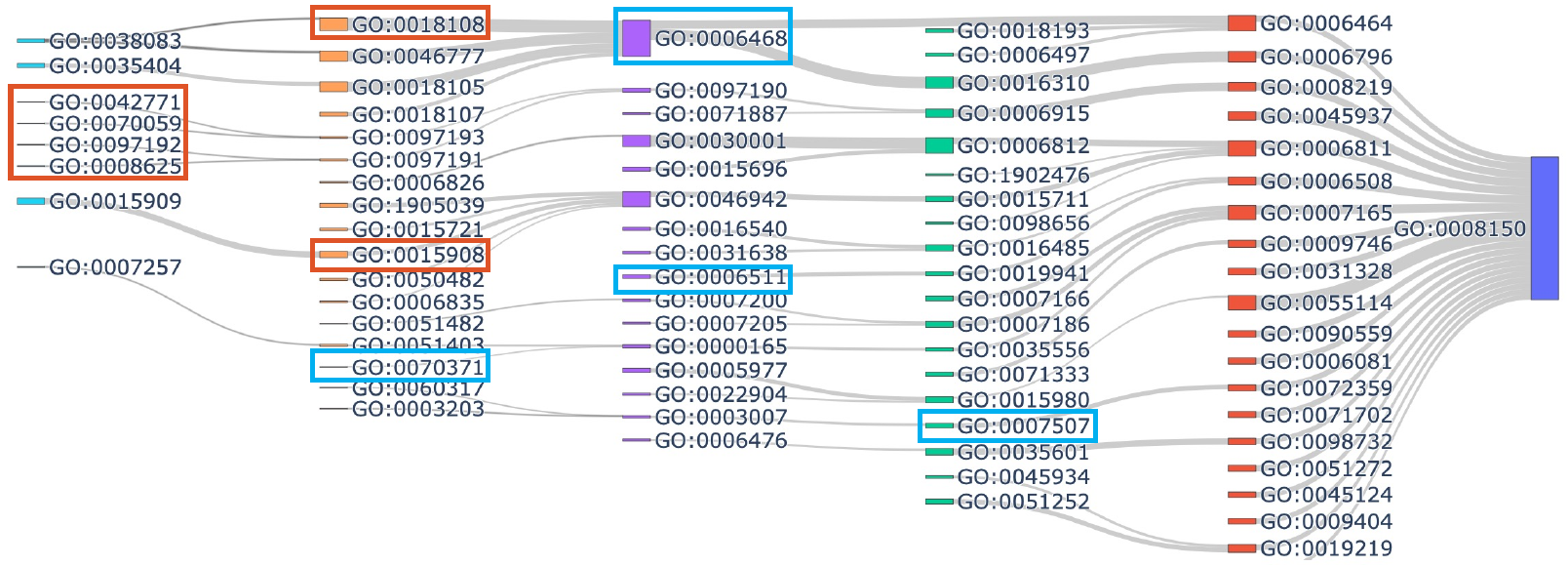}}\\\vspace{-1ex}
 \subfloat[\label{fig:sankey_DOX_BRCA}]{\includegraphics[width=.92\textwidth]{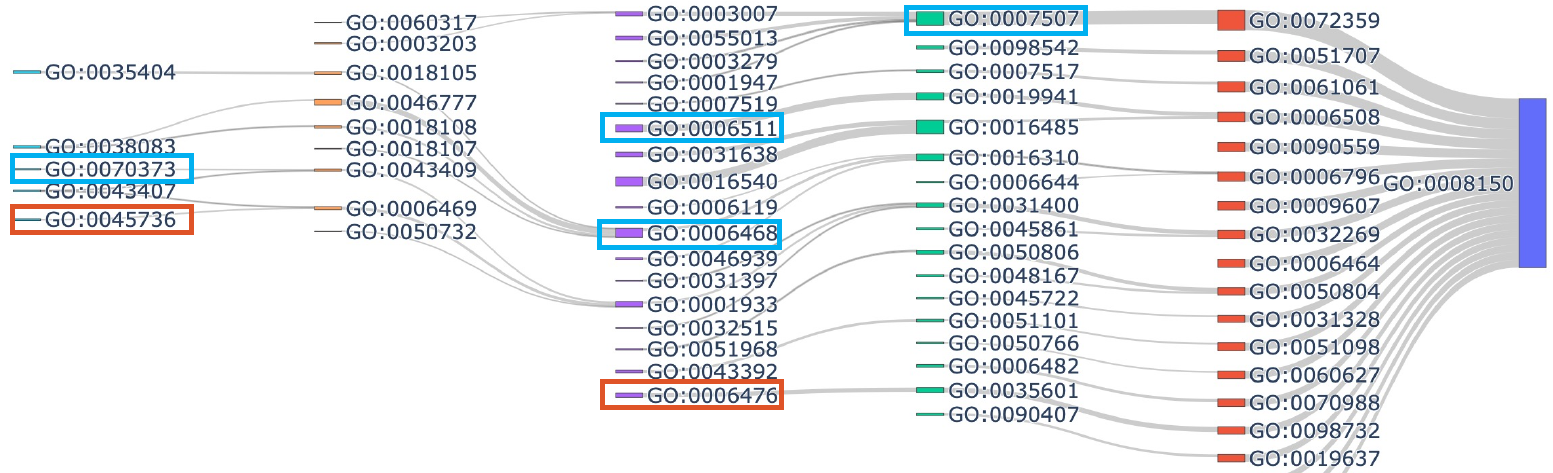}}\\
    \caption{Hierarchical explanation of VETE for: (a) Docetaxel-OV pair; (b) Docetaxel-BRCA pair. Blue squares highlight selected shared GO terms between the two pairs, while red squares highlight selected unique GO terms.}
    \label{fig:drug_cell_explain}
\end{figure*}

Given the efficiency of VETE in encoding cell line transcriptomics data, we further incorporated drug embedding and trained VETE as a regression model to predict drug responses for drug-cell pairs, where the response is the Area Under the dose-response Curve (AUC). The model was trained with Adam. We employed Spearman's rank correlation coefficient as the performance metrics and compared the model performance with MLP, XGBoost, and Random Forest. For MLP, XGBoost, and Random Forest, gene expression and drug molecular structure embeddings are concatenated together as the model input.

Experiment results are shown in Figure \ref{fig:drug_cell}. VETE achieves an overall correlation coefficient of 0.725 on the testing set. We analyzed the prediction accuracy across different actual AUC intervals (Figure \ref{fig:drugcell_perf_box}), revealing that VETE performs well across varying treatment sensitivities. However, it slightly overestimates low sensitivity samples due to lack of data points. Samples with AUC less than 0.2 account for only 0.14\% of the total samples (157 out of 114,644 drug-cell pairs) and are omitted in evaluation. Comparatively, VETE, XGBoost and Random Forest outperform MLP in this study, with VETE having the highest performance among the four models (Figure \ref{fig:drugcell_compare}). Additionally, we investigate the prediction performance for individual drugs and identify 10 drugs associated with the highest prediction performance, where model performance varies across different types of drugs (Figure \ref{fig:indi_drug}). Interestingly, the model performs the best for vinca alkaloids drugs (vinblastine, vincristine, and vinorelbine) and YK-4-279, which have been shown to exhibit synergy with vinca alkaloids in a previous study \citep{zollner2017inhibition}.

We used ovarian cancer (OV) and breast cancer (BRCA) samples treated with Docetaxel as the examples to demonstrate VETE’s interpretability in providing insights into drug-cell interactions. Docetaxel,  a type of taxane drugs that interferes with microtubules, is commonly used to treat breast cancer, ovarian cancer, non-small cell lung cancer, etc. \citep{lyseng2005docetaxel}. We selected 56 Docetaxel-OV samples and 155 Docetaxel-BRCA samples for the analysis.

Several GO terms are shared across the two cancer types (highlighted with blue squares in Figure \ref{fig:drug_cell_explain}). For instance, Docetaxel is known to interfere with microtubules to suppress cell proliferation by decreasing phosphorylation \citep{lyseng2005docetaxel}, where protein phosphorylation (GO:0001933 and GO:0006468) are identified in both plots. Similarly, ERK1/2 (GO:0070371/70373) can also be influenced by microtubule-targeting agents and are shared in the plots \citep{stone2000microtubule}.
We observed that GO terms and paths related to heart development (GO:0007507) are considered to influence the sensitivity of OV and BRCA cells to Docetaxel. Docetaxel is known for its cardiotoxicity \citep{shimoyama2001docetaxel}, and several studies and cases have reported the potential heart failure issues associated with treating BRCA with Docetaxel that requires additional care in the treatment \citep{mackey2013adjuvant, mase2023case}. 
Studies suggest that ubiquitination  is involved in the regulation of metabolic reprogramming in cancer cells \citep{shi2010ubiquitin, deng2020role, han2023ubiquitin}. For example, a recent study in prostate cancer report that the deubiquitinating enzyme ubiquitin-specific protease 33 inhibits Docetaxel-induced apoptosis of prostate cancer cells \citep{guo2020deubiquitinating}. The results in Figure \ref{fig:drug_cell_explain} imply that enzyme deubiquitylation may play a similar role in breast cancer and ovarian cancer (GO:0006511/0031397).

Besides the shared GO terms, unique GO terms exist that are specific to each cancer type (highlighted with red squares in Figure \ref{fig:drug_cell_explain}).
Figure \ref{fig:sankey_DOX_OV} the Sankey plot for Docetaxel-OV samples, where several GO terms align with existing literature findings. For example, previous studies suggest the association of tyrosine phosphorylation (GO:0018108) and ovarian cancer \citep{song2019proteome} and that Docetaxel administration can reduce certain types of phosphorylation in castrate-resistant prostate cancer \citep{lee2014phosphoproteomic}. However, the association of Docetaxel and ovarian cancer through phosphorylation is rarely studied and require further investigation. Numerous apoptotic signaling pathways (e.g., GO:0042771, GO:0070059, GO:0097192, GO:0008625) identified in VETE have been carefully studied in relation to Docetaxel sensitivity \citep{liu2013functional, lee2023Docetaxel} and ovarian cancer \citep{zhao2017hesperidin, nie2022herpud1}. Fatty acid transport (GO:0015908) is also associated with the increased incidence and aggressiveness of ovarian cancer \citep{yoon2022fatty}, and previous studies have shown that Docetaxel and fatty acid binding protein produce synergistic inhibition of prostate cancer \citep{carbonetti2020docetaxel}. 

\begin{figure}[!tb]
    \centering
    \includegraphics[width=1\linewidth]{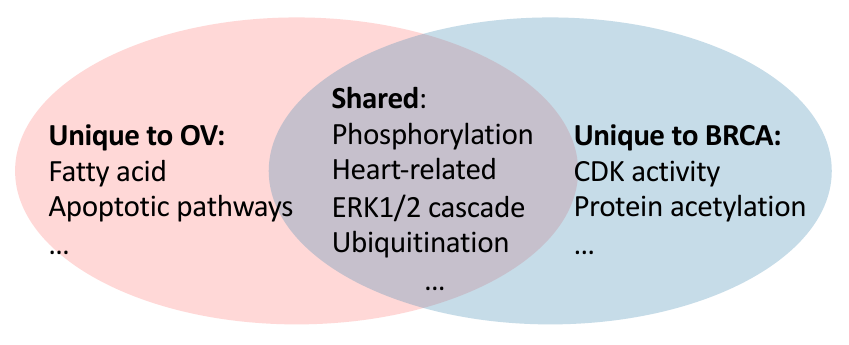}\vspace{-3ex}
    \caption{Shared and unique GO terms under OV and BRCA.}\vspace{-3ex}
    \label{fig:shared_go_terms}
\end{figure}

Similar consistency is observed in Figure \ref{fig:sankey_DOX_BRCA}. 
Cyclin-dependent protein serine/threonine kinase activity (CDK)(GO:0045736) is identified as related to the drug response of breast cancer samples, where the use of CDK4/6 inhibitors for breast cancer treatment has drawn significant attention in recent years \citep{shah2018cdk4, sofi2022cyclin, wang2024recent}. Although Docetaxel has not been directly linked to CDK4/6, studies have found that pharmacological inhibition of CDK4/6 yields a cooperative cytostatic effect when combined with Docetaxel \citep{kumarasamy2020chemotherapy, de2022phase}.
Previous studies also revealed that inhibition of ERK1/2 can reverse Docetaxel resistance in MCF-7 spheroid culture, a human breast adenocarcinoma cell, which is also depicted in the Sankey plot (GO:0070373) \citep{jeong2010role}. Atypically acetylated proteins (GO:0006476) have been shown to promote breast cancer metastasis and proliferation \citep{riolo2012histone, chang2014acetylation}, which may also cause Docetaxel resistance \citep{li2020tgf}. 

A summary of the selected shared and unique GO terms in the two Sankey plots are summarized in Figure \ref{fig:shared_go_terms}. Although transcriptomics submodel relies solely on gene expression and biological processes, and the gene embedding is concatenated with drug embedding in the final layers of VETE's encoder, several drug-related GO terms are identified in the pruned Sankey plots. This finding highlights VETE’s potential in effectively interpreting drug response predictions.

\section{Discussion}

In this work, we propose VETE, an interpretable variational neural network designed to mirror biological processes for encoding cancer profiles and predicting drug responses. To further enhance the explainability of the fitted model, we developed a graph-based post-hoc explanation method that assists in searching and pruning the biological graph from VETE. As the quality of explanations is sensitive to the model's fit, we employ large-scale hyperparameter optimization using a centralized Bayesian optimization framework. Our final results show consistency with existing biological findings. We anticipate that these proposed methodologies will advance precision oncology by uncovering novel molecular mechanisms of drug responses and inspiring new treatment strategies.

Despite its strengths, the proposed method has limitations. For example, while VETE can effectively identify important GO terms and paths, the results can be dominated by the most influential GO terms, which are already well-studied and well-analyzed in the literature. Alternative searching and pruning algorithms may be needed to discover previously unknown and relatively less influential GO terms. The VIB framework makes parametric and mean field assumptions on the posterior distribution. Models have been proposed to relax these assumptions for more flexible model fitting \citep{shao2024nonparametric, tran2015copula}, and investigating these tools could further improve model performance. 
Selecting appropriate baselines is critical due to their substantial impact on  explanation results. Additional works are needed to explore the robustness of the proposed method to baselines and develop strategies for automated and data-adaptive baseline selection.
Lastly, additional experiments are needed for assessing the method's generalizability and transferability using Patient-Derived Xenograft (PDX) datasets, such as the MMHCdb \citep{begley2023mouse}.

\section*{Acknowledgements}

This research has received support from the following funding sources: federal funding through the NCI-DOE collaboration established by the U.S. Department of Energy (DOE) and the National Cancer Institute (NCI) of the National Institutes of Health, Cancer Moonshot Task Order No. 75N91019F00134, Frederick National Laboratory for Cancer Research contract 75N91019D00024, and the generosity of Eric and Wendy Schmidt through the Eric and Wendy Schmidt Fund for AI Research and Innovation at Mayo Clinic.

\bibliography{main}
\bibliographystyle{icml2024}

\newpage
\appendix
\onecolumn
\section{VETE for Cancer Type Prediction with TCGA}

\begin{figure}[!htb]
    \centering
	\subfloat[\label{fig:tcga_confusion}]{\includegraphics[width=.42\textwidth]{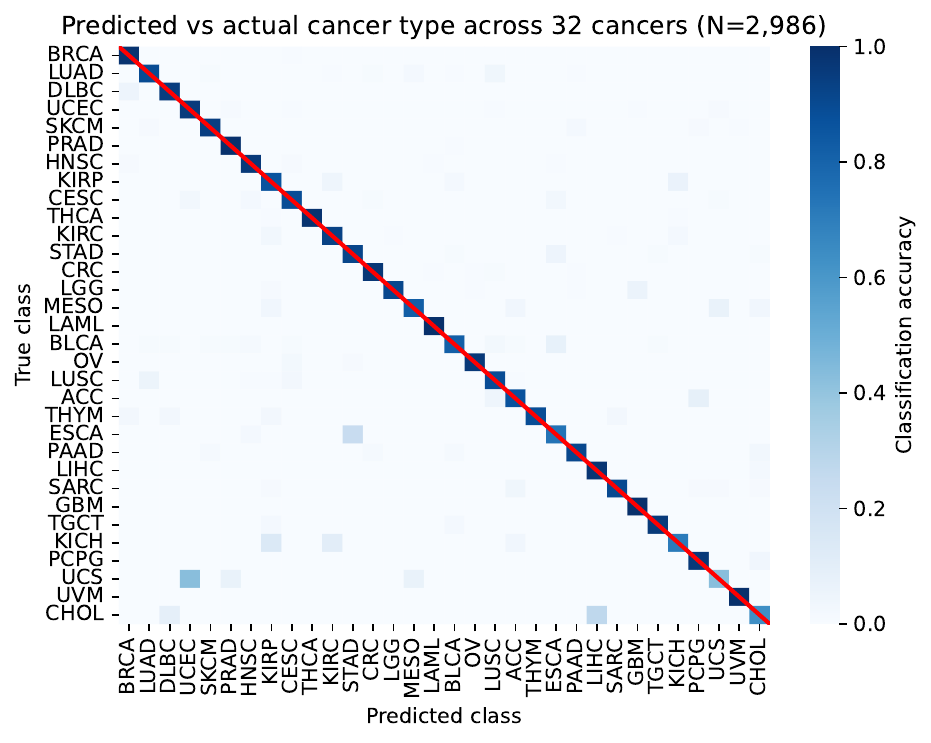}}
    \subfloat[\label{fig:tcga_tsne}]{\includegraphics[width=.325\textwidth]{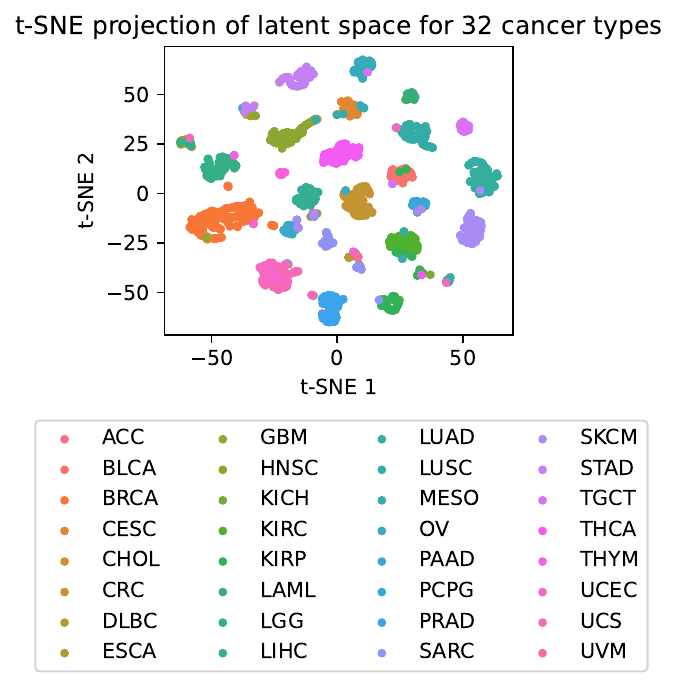}}
    \subfloat[\label{fig:tcga_perf}]{\includegraphics[width=.25\textwidth]{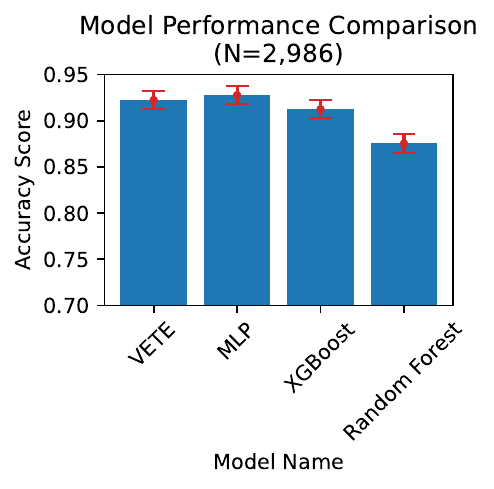}}\\
    \caption{TCGA multiclass classification results: (a) Confusion matrix for multiclass classification perfor-mance across all 33 cancer types. Rows represent the true classes, and columns the predicted clas-ses; (b) Visualization of latent space of VETE for testing set; (c) Performance comparison of VETE again MLP, XGBoost, and Random Forest.}
    \label{fig:TCGA_clf_res}
\end{figure}

\begin{figure}[!htb]
    \centering
	\includegraphics[width=.9\textwidth]{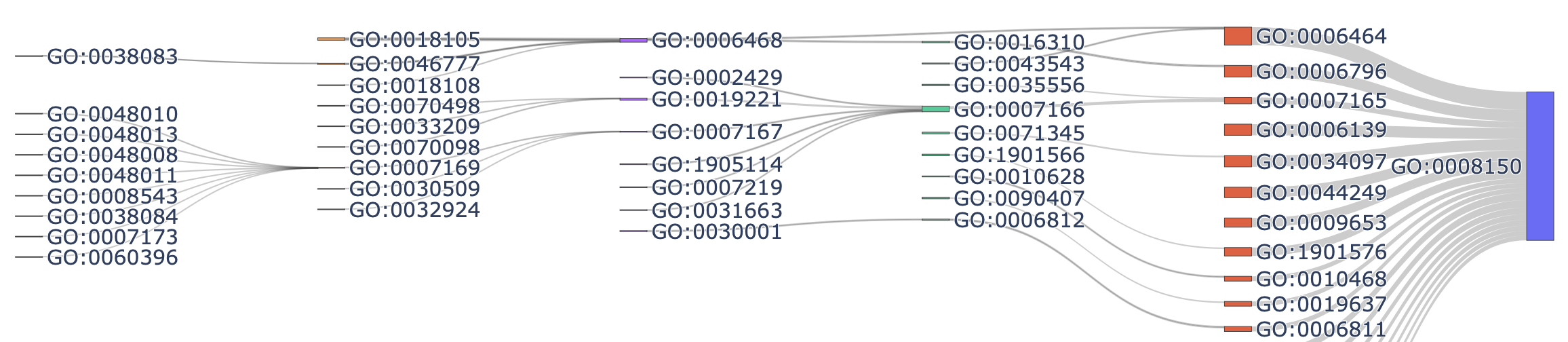}
    \caption{Hierarchical explanation of VETE for OV under TCGA.}
    \label{fig:sankey_tcga}
\end{figure}

We conducted a multiclass classification experiment using TCGA to validate the transcriptomics submodel's effectiveness for gene expression encoding. We used processed RNAseq data from  PanCanAtlas TCGA organized through  the BioBombe study \citep{weinstein_cancer_2013, way_compressing_2020}. This dataset comprised transcriptomics samples from 11,060 diagnosed patient tumors, covering 32 different cancer types and 16,148 genes.  We used the intersection of the selected 3,008 genes from CCLE in the main text and genes in TCGA, resulting in 2,702 genes for our experiments and analysis.  The dataset was then divided into a training set (6,968 samples) and a testing set (2,986 samples). The objective was to classify each sample into one of the 32 primary cancer types. For this task, we implemented a one-hidden-layer MLP with a singular output neuron as the decoder. This simple decoder design aimed to improve the encoder's learning of sample embeddings.  Cross-entropy was used as the loss function. We report the prediction accuracy for individual classes and compare the model performance against popular machine learning models: Random Forest, XGBoost, and fully connected MLP with the same structure as the decoder of VETE. 

The classification results are presented in Figure \ref{fig:TCGA_clf_res}. Figure \ref{fig:tcga_confusion} depicts the confusion matrix of multiclass classification, where the rows represent the true classes, and the columns represent the predicted classes. Depth of the $ij$-th square’s color implies the percentage of samples belonging to class $i$ and predicted as class $j$. Overall, VETE achieves good prediction performance across most cancer types. Prediction is not as good for READ (rectal adenocarcinoma) and UCS (uterine carcinosarcoma), where samples can be misclassified as COAD (colon adenocarcinoma) and UCEC (uterine corpus endometrial carcinoma). In Figure \ref{fig:tcga_tsne}, we project the latent space of testing samples from VETE to a two-dimensional space using t-SNE [40], where samples from different cancer types are well-separated, encouraged by the classification task. Performance comparison shows the advantage of VETE over the other three methods (Figure \ref{fig:tcga_perf}).

We identify multiple hierarchical GO paths that positively influence the classification results for BRCA samples (Figure \ref{fig:sankey_tcga}). It is interesting to note that platelet-derived growth factor receptor signaling pathway (GO:0048008) is identified as a positive contribution towards a sample being classified as ovarian cancer, which coincides with previous biological study \citep{matsuo2014platelet}. Another interesting identified GO term is fibroblast growth factor receptor signaling pathway (GO:0008543), where previous studies suggest its association with ovarian cancer risk \citep{meng2014genetic}. Besides identifying important low level GO terms, the hierarchical explanation graph also depicts the path through which the low-level GO terms influence the final model prediction.


\end{document}